# Cytogenetic, Hematobiochemical, and Histopathological Assessment of Albino Rats (*Rattus norvegicus*) Fed on Gluten Extracts

Tajudeen Yahaya[1*], Esther Oladele[2], Ufuoma Shemishere[3], Daniel Anyebe[4], Haliru Abdullahi[5], Maryam Lawal[6], and Rufa'i Ahmad[7]

[1]Department of Biological Sciences, Federal University Birnin Kebbi, Nigeria
[2]Biology Unit, Distance Learning Institute, University of Lagos, Nigeria
[3]Department of Biochemistry and Molecular Biology, Federal University Birnin Kebbi, Nigeria
[4]Department of Biochemistry and Molecular Biology, Federal University Birnin Kebbi, Nigeria
[5]Department of Biological Sciences, Federal University Birnin Kebbi, Nigeria
[6]Department of Biological Sciences, Federal University Birnin Kebbi, Nigeria
[7]Department of Biological Sciences, Federal University Birnin Kebbi, Nigeria

**Abstract**

**Background:** Literature shows that most of the information on the toxicity of gluten is generated from survey and observational studies, resulting in inconsistent outcomes and a decrease in the acceptability of gluten-rich foods. To determine gluten's safety, an in-depth *in vitro* and in *vivo* toxicological examination is required. This enables scientists to come up with ameliorative strategies if it turns out to have side effects, and consumers' trust can be restored.

**Objectives:** The objective of this study was to assess the toxicity of gluten extracts on albino rats (*Rattus norvegicus*).

**Materials and Methods:** twenty-four rats were randomly selected and divided into four groups, each comprising six rats. Group 1 (control) rats were fed on pellet feeds and groups 2, 3, and 4 were fed on daily dosages of 0.5, 1.0, and 1.5 g gluten extracts, respectively. The rats' body weights and reactions were observed for 90 days before blood samples were collected for hematobiochemical and micronucleus tests. Histopathological examinations of the liver and kidneys were also performed.

**Results:** There was no difference ($P > 0.05$) in body weight, blood glucose level, or micronuclei between the control and treated rats. The lymphocytes, alkaline phosphatase, alanine transaminase, total protein, and calcium ions of the test rats were all significantly ($P < 0.05$) altered but remained within the normal ranges. Other hematobiochemical parameters, including packed cell volume, hemoglobin, white and red blood cells, aspartate transaminase, albumin, sodium ions, potassium ions, chloride ions, and urea, revealed no marked changes. The treated rats' livers and kidneys showed no histopathological changes.

**Conclusion:** Gluten had no adverse effects on body weight, blood glucose level, genetic constitution, or histopathology in the treated rats. However, it altered hematobiochemical parameters, particularly the lymphocytes, alkaline phosphatase, alanine transaminase, total protein, and calcium ions. This suggests that prolonged consumption of gluten-containing diets can affect hematobiochemical parameters.

**Keywords:** Alkaline phosphatase; Gluten; Lymphocytes; Micronuclei; Sodium ion

## 1. Introduction

Gluten is a mixture of seed storage proteins present in grains such as wheat, rye, barley, and oats (Biesiekierski, 2017). It is composed mainly of gliadin (prolamin) and glutenin (glutelin), which are collectively called prolamins (Biesiekierski, 2017). Gliadin is soluble in alcohol, but glutenin is insoluble and plays an important role in baking (Yeonjae *et al*., 2018). When properly hydrated and mixed, gluten in flour forms a three-dimensional protein network that is used in baking to create viscoelastic dough and bread products (Laszkowska *et al*., 2018; Niland and Cash, 2018). Gluten is also used as a food additive in desserts, pastas, and sauces, among other things (Laszkowska *et al*., 2018; Sfgates, 2018).

Cereals containing gluten such as wheat, barley, and oats, are staple foods all over the world (Shewry, 2009). Wheat, in particular, is the main source of food for about 40% of the people in the world (Giraldo *et al*., 2019; Grote *et al*., 2021). In some parts of the world, people prefer wheat to rice (Yeonjae *et al*., 2018). In Nigeria, wheat is the third most consumed grain after corn and rice (Lyddon, 2022). Global annual wheat production is about 750 million tons (OECD/FAO, 2019). In 2019, world cereal production was over 2700 million tons, of which wheat, maize, and barley accounted for the majority of the harvests (FAO, 2019).

Wheat and other cereals have several dietary components that can help alleviate undernutrition, micronutrient malnutrition, and overnutrition (Poole *et al.*, 2020). Cereals are important sources of carbohydrates, protein, vitamins, and phytochemicals (Shewry and Hey, 2015). There is a proven link between the fiber content of cereals and reduced risks of heart disease, diabetes, and cancer (Shewry and Hey, 2015).

Unfortunately, there is an age-old controversy concerning the safety of gluten. According to Navarro *et al.* (2017), Giménez *et al.* (2019), Tanveer and Ahmed (2019), and Antvorskov *et al.* (2018), the gliadins and glutenins in gluten are immunogenic and can cause allergies as well as intestinal and coeliac disease and type 1 diabetes in susceptible people. In contrast, some studies, including Neyrinck *et al.* (2012), Lebwohl *et al.* (2017), and Yahaya *et al.* (2020a), have reported the non-toxicity of gluten. In addition, a few studies, including Sapone *et al.* (2016) and Freeman (2018), cautioned that nutritional plans lacking gluten-rich foods may cause malnutrition and exposure to heavy metals. These contrasting results show that the safety of gluten has not been ascertained, and this is fast eroding the confidence of consumers.

In Nigeria, over 100 thousand cases of gluten intolerance are recorded annually, which has resulted in the reduced acceptability of gluten containing foods such as bread and a popular local food called amala (NIFST, 2023). However, there is scarcity of documented information in the country to suggest that consumption of wheat causes gluten intolerance. Thus, in-depth toxicological examinations (both *in vivo* and *in vitro*) of gluten are imperative in order to create ameliorative strategies if it is discovered to have harmful effects. This will go a long way towards restoring consumers' trust in gluten-containing diets and, by extension, boosting the production and marketing of gluten-containing cereals. This study, therefore, was conducted to investigate the effects of gluten on body weight, hematobiochemical parameters, cytogenetics, and histopathology in albino rats (*Rattus norvegicus*). Hematobiochemical parameters are sensitive to various stressors and thus are good indicators of health (Abdullah *et al.*, 2022). Cytogenetic aberrations are a biomarker for early effects that indicate the cell or organism has undergone chromosomal damage as a result of exposure to an external mutagenic or carcinogenic agent (García Sagredo, 2011). Histopathology is an important part of animal model studies because it gives morphologic context to *in vivo*, molecular, and biochemical data (Knoblaugh *et al.*, 2018).

## 2. Materials and Methods

### 2.1 Description of Study Site

This study was carried out in Birnin Kebbi, the capital city of Kebbi State, northwestern Nigeria. The state can be located on latitude 12°21'N and longitude 4°36'E (Ajala *et al.*, 2019). The neigbouring states to Kebbi State are Katsina and Zamfara States on the west, Sokoto State on the northern part, and Niger State on the south. The eastern parts of the state are bordered by the Niger and Benin Republics. The indigenous people of the state comprises of Hausa, Fulani, and Zuru and has setlers like Yoruba, Igbo, Nupe, and some other tribes. The indigenous people of the state are predominantly farmers and animal breeders and are known in the country for cereal production. Wheat, corn, and rice are among the cereals that are planted and consumed in large quantities by the residents of the state. Considering the controversy surrounding the safety of gluten-rich foods such as wheat, it is imperative to evaluate the toxicity of gluten so as to advise consumers in the state appropriately.

### 2.2. Source and Management of Experimental Animals

The guidelines for the care and use of animals in research and teaching issued by the Animal Ethics Committee of the Federal University Birnin Kebbi, Nigeria, were followed in this study. Twenty-four (24) mixed-sex albino rats (*Rattus norvegicus*) aged 55 days, with a mean weight of 195 ± 2 g, were purchased from the Department of Biological Sciences, Federal University Birnin Kebbi, in October 2021. The rats were kept in metal cages and had free access to water and commercial pellet rat chow from the Vital Feed Industry, Lagos. Normal weather conditions were maintained at an ambient temperature of 28 ± 4 °C. Before starting the experiment, the rats were given a two-week acclimatization period.

### 2.3. Source and Extraction of Gluten

The procedures of Bathula *et al.* (2018) were followed to extract gluten from wheat purchased in June 2021 from Birnin Kebbi Central Market. The wheat grains were milled into powder using a mortar and pestle. One hundred grams (100 g) of the wheat flour were scooped into 70 ml of water and left to stand for one hour. The dough was washed under running water to remove starch and impurities in order to obtain pure gluten, characterized by its viscoelasticity. The extracted gluten was broken into small fragments with a mortal and pestle, air-dried, ground, and kept in a desiccator prior to use.

### 2.4. Treatments and Experimental Design

The twenty-four (24) rats were randomly divided into four groups of six (6) each. The control rats in group one were fed only pellet feed, while groups two, three, and four were daily administered (orally) 0.5, 1.0, and 1.5 g of gluten extract, respectively. The rats' body weights, blood glucose levels, and overall reactions were tracked for 90 days before blood samples were obtained for hematological, hepatic, and renal functions and cytogenetic tests. Histopathological examinations of the rats' liver and kidney tissues were also performed.

### 2.5. Blood and Tissue Sample Collection

Office pins were used to hold each rat to a work bench. A 5-ml syringe and a 20-gauge needle were used to draw 2.5 ml of blood through a cardiac puncture into ethylenediamine tetraacetic acid (EDTA) bottles. The rats were thereafter sacrificed by cervical dislocation to collect their livers and kidneys for histopathological examinations. The liver and kidneys are selected because they are the primary target of toxins as they process all substances ingested, especially the liver.

### 2.6. Body Weight Measurement

A digital weighing balance [KERN (EMB), Germany] was used to measure the rats' body weights at the beginning of the experiment. Throughout the experiment, measurements were taken in the morning at five-day intervals.

### 2.7. Blood Glucose Level Measurement

The fasting blood glucose of the rats was measured using the Fantastik-Accu Glucose Meter (IVD version 180705-1), as described by Yahaya (2017). A drop of blood was drawn from each rat and placed on a test strip. The strip was put into the glucose meter, which displayed the reading in milligrams per deciliter (mg dL$^{-1}$).

### 2.8. Hematological Tests

The hematological tests were carried out following the procedures used by Yahaya *et al.* (2020b). A blood analyzer (Sysmex auto-analyzer model XP-300) was used to measure blood parameters like the packed cell volume (PCV), hemoglobin (Hb), white blood cells (WBC), red blood cells (RBC), and lymphocytes (LYM).

### 2.9. Hepatic Function Tests

The hepatic (liver) function tests were conducted as detailed by Yahaya *et al.* (2020b). The blood samples were left to clot at room temperature, after which the clots were removed by centrifuging the sample for 10 minutes at 2000 x g. The supernatant (serum) obtained was poured into a pre-washed polypropylene tube and used to determine the liver enzymes, including alanine transaminase (ALT), aspartate transaminase (AST), alkaline phosphatase (ALP), albumin (ALB), and proteins. Alanine transaminase, AST, and ALP were estimated by ultraviolet, colorimetric, and spectrophotometric methods at 546 nm, respectively, while the biuret method and bichromatic digital endpoints were used to estimate total proteins and ALB, respectively.

### 2.10. Renal Function Tests

The renal (kidney) function tests were determined from the blood serum, as outlined by Yahaya *et al.* (2019). The serum sodium (Na$^+$), potassium (K$^+$), chlorine (Cl$^{-1}$), calcium (Ca$^{2+}$), and urea levels were measured using an autoanalyzer (SKU: SM100).

### 2.11. Micronucleus Assay

The micronucleus test was conducted in accordance with the OECD Guidelines (2007). Two (2) mL of saline solution were used to collect bone marrow from the femur of each rat. The bone marrow was centrifugated for 7 minutes after which 10 μL of bone marrow or blood was placed on three slides per rat. The slides were fixed in absolute methanol for 15 minutes, airdried, and stained with 10% Giemsa. The slides were washed with distilled water and allowed to dry. A digital microscope was used to view the slides for nuclear abnormalities (NAs) at a magnification of 1000. A total of 3000 erythrocytes were scored per rat (1000 erythrocytes per slide).

### 2.12. Histopathological Analysis

The histopathogical examinations were done as conducted by Yahaya *et al.* (2020a). Liver and kidney tissue samples (5 mm thick) were fixed in 10% formal saline for 8 hours. After dehydration in different alcohol concentrations (100%, 85%, and 65%), the tissues were hardened in molten wax. Remaining alcohol and wax were removed using various xylene concentrations. The embedded tissues were attached to a wooden block and sections were detached with a 20% alcohol solution. Following warm water treatment, the sections adhered to microscopic slides. The slides were stained with hematoxylin and eosin for examination under a light microscope.

### 2.13. Statistical data analysis

All analyses were conducted using the Statistical Package for Social Science (SPSS) version 21 for Windows. For the micronucleus test, data were presented as mean ± standard error, and for the blood glucose level, body weight, hematological, hepatic, and renal function tests, data were presented as mean ± standard deviation. The differences between the treated and control groups were determined using analysis of variance (ANOVA). A probability level at P ≤ 0.05 was chosen as the statistical significance level.

## 3. Results

### 3.1. Effects of Gluten on Body Weight, Blood Glucose Levels, and Hematological Parameters

The body weights, blood glucose levels, and hematological parameters of rats administered gluten extracts are shown in Table 1. There were no significant (P ≥ 0.05) differences between the treated and control groups regarding body weights and blood glucose levels. With the exception of lymphocytes (LYM), there were no significant (P ≥ 0.05) differences between the blood parameters of the control and treated groups (see also supplementary Tables 1–3).

Table 1. Body weights, blood glucose levels, and hematological parameters of rats treated with gluten extracts.

| Conc (g) | Body weight (g) | | Blood glucose (mmol L$^{-1}$) | | Haematological Parameters | | | | |
|---|---|---|---|---|---|---|---|---|---|
| | Initial | Final | Initial | Final | PCV (L L$^{-1}$) | Hb (g dL$^{-1}$) | WBC (mc mm$^{-3}$) | RBC (mc mm$^{-3}$) | LYM (c μL$^{-1}$) |
| Control | 199.33±6.86 | 207.41±3.86 | 5.05±0.08 | 5.02±0.03 | 0.26 ± 0.01 | 11.22 ± 2.19 | 14.23 ± 6.30 | 5.20 ± 0.98 | 63.32 ± 5.41 |
| 0.5 | 198.33±6.09 | 210.51±5.41 | 5.22±0.12 | 5.13±0.07 | 0.27 ± 0.03 | 12.87 ± 0.57 | 14.57 ± 6.17 | 5.76 ± 0.51 | 71.62 ± 2.24* |
| 1.0 | 197.00±3.29 | 209.32±4.26 | 5.17±0.05 | 5.22±0.09 | 0.27 ± 0.03 | 12.08 ± 0.98 | 14.00 ± 4.08 | 5.63 ± 0.56 | 72.98 ± 3.95* |
| 1.5 | 198.15±3.54 | 208.43±3.71 | 5.22±0.17 | 5.14±0.08 | 0.26 ± 0.02 | 11.15 ± 1.53 | 13.72 ± 2.84 | 5.57 ± 0.67 | 72.18 ± 3.17* |

Note: *The values were expressed as mean ± standard deviation (n = 6); values with an asterisk (*) in the row are statistically different from the control at P ≤ 0.05 (ANOVA); PCV Stands for packed cell volume; Hb represents hemoglobin; WBC indicates white blood cells; RBC denotes red blood cells; and LYM is short for lymphocytes.*

Table 2. Hepatic and renal function parameters of rats treated with gluten extracts.

| Conc (g) | Liver Function Parameters | | | | | Renal Function Parameters | | | | |
|---|---|---|---|---|---|---|---|---|---|---|
| | ALP (U L$^{-1}$) | AST (U L$^{-1}$) | ALT (U L$^{-1}$) | TP (g dL$^{-1}$) | ALB (g dL$^{-1}$) | Na$^+$ (mEq $^{-1}$) | K$^+$ (mmol $^{-1}$) | Cl$^-$ (mEq $^{-1}$) | Urea (mg dl$^{-1}$) | Ca$^{2+}$ (mmol $^{-1}$) |
| Control (0) | 21.00±1.00 | 11.71±1.4 | 13.00±1.02 | 7.00±1.00 | 4.07±1.3 | 140.17±1.2 | 4.15±0.2 | 102± 1.4 | 3.05± 0.2 | 76.5 ± 6.5 |
| 0.5 | 17.67±2.52* | 12.69±2.52 | 07.67±3.06* | 8.43±2.41* | 4.13±1.15 | 140.17 ± 1.47 | 4.05 ± 0.34 | 102.83 ± 1.47 | 3.03 ± 0.53 | 71.67 ± 8.64 |
| 1.0 | 15.33±2.43* | 11.65±1.53 | 07.33±1.53* | 7.80±1.72* | 4.06±1.23 | 136.17 ± 11.20 | 4.05 ± 0.33 | 103.00 ± 1.41 | 2.93 ± 0.47 | 82.50 ± 8.12 |
| 1.5 | 15.81±2.61* | 11.81±1.31 | 07.45±1.01* | 8.02±1.20* | 4.10±1.11 | 142.33 ± 2.07 | 4.07 ± 0.23 | 102.83 ± 1.17 | 3.02 ± 0.37 | 66.50 ± 7.87* |

Note: *The values were expressed as mean ± standard deviation (n = 6); values with an asterisk (*) in the columns are statistically different from the control at p ≤ 0.05 (ANOVA); ALP = Alkaline phosphatase; AST = Aspartate transaminase; ALT = Alanine transaminase; TP = Total protein; ALB = Albumin; Na$^+$ = Sodium ion; K$^+$ = Potassium ion; Cl$^-$ = Chloride ion; Ca$^{2+}$ = Calcium.*

### 3.2. Effects of Gluten on Hepatic and Renal Functions

Table 2 reveals the hepatic and renal function parameters of gluten-fed and control rats. Regarding the hepatic function, the treated groups' alkaline phosphatase (ALP) and alanine transaminase (ALT) levels were significantly ($P \leq 0.05$) lower; total protein (TP) levels were significantly ($P \leq 0.05$) higher; and aspartate transaminase (AST) and albumin (ALB) levels were unaffected ($P \geq 0.05$). There were no significant ($P \geq 0.05$) changes in the renal functions of the treated rats, except for the $Ca^{2+}$ of the rats treated with 1.5 g of the extract. For more details, see supplementary Tables 4 and 5.

### 3.3. Genotoxic Effects of Gluten

Table 3 and supplementary Table 6 depict the micronuclei and nuclear abnormalities observed in the gluten-treated and control rats. There were no significant differences ($P > 0.05$) between the rats that were treated and the rats that were not. All groups had micronuclei and nuclear abnormalities (Figure 1a–d).

Table 3. Micronuclei and nuclear aberrations in the rats treated with gluten extracts.

| Conc (g) | Total cell examined | micronucleus | Nuclear aberration |
|---|---|---|---|
| Control | 3000 | 5.33±0.84 | 6.50±3.21 |
| 0.5 | 3000 | 5.33±0.88 | 6.44±3.20 |
| 1.0 | 3000 | 6.67±0.67 | 6.33±5.96 |
| 1.5 | 3000 | 6.33±1.45 | 6.40±4.09 |

Note: *The values were expressed as mean ± standard errors (n = 6); values without an asterisk (\*) in the columns are statistically not different from the control at $P \leq 0.05$ (ANOVA).*

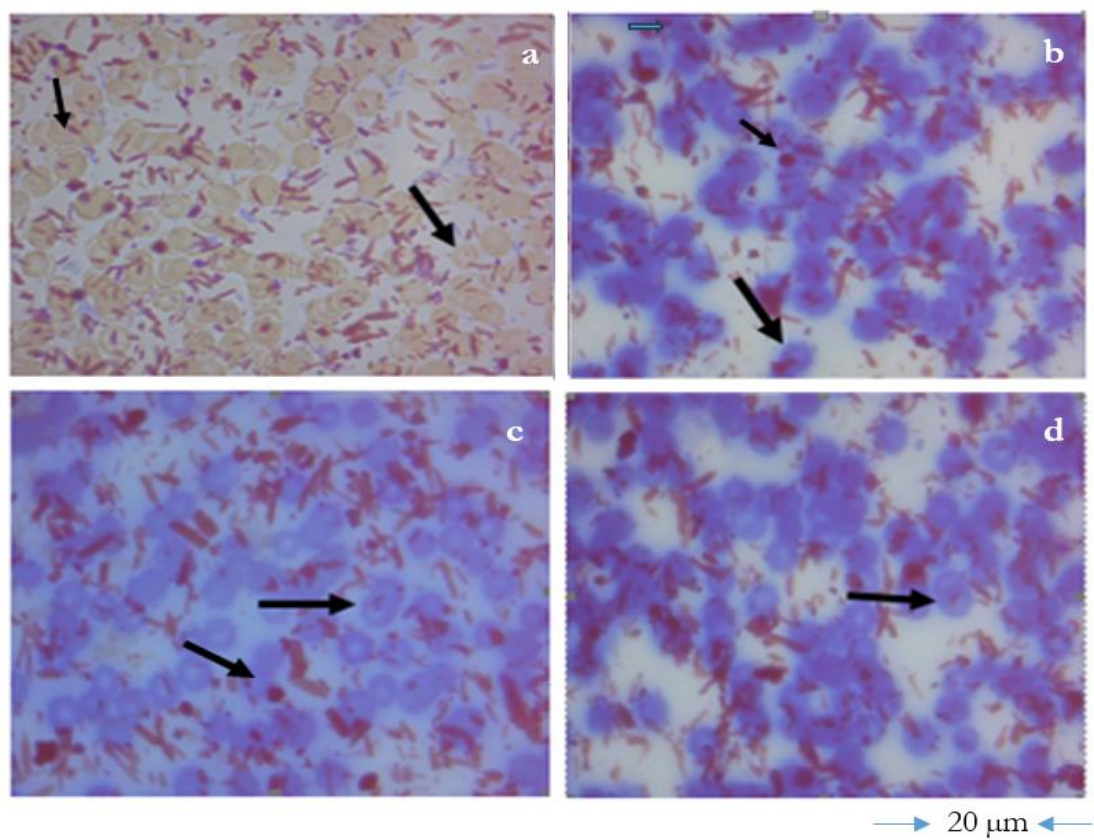

Figure 1. Photomicrograph of blood cells of rats treated with gluten extracts. Control rats showing micronuclei (a), rats fed 0.5 g of gluten extract showing micronuclei (b), rats fed 1.0 g of gluten extract showing micronuclei (c), and rats fed 1.5 g of gluten extract showing micronuclei (d) x 100.

### 3.4. Histopathological Effects of Gluten

Figure 2a–h show the effects of gluten extracts on the kidney and liver tissues of the treated rats. Compared to the kidneys and livers of the control rats, the kidneys and livers of the treated rats did not show any changes in histology.

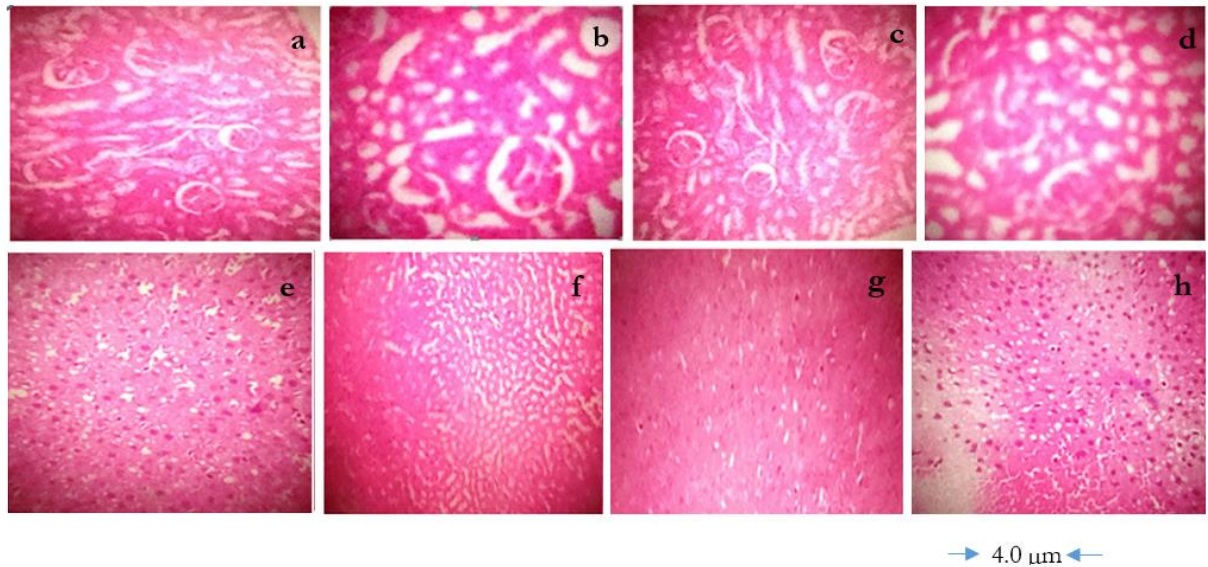

Figure 2. Photomicrographs of kidney and liver tissues of rats fed gluten extracts (x100). Kidneys of control rats showing normal glomeruli (a), kidneys of rats fed 0.5 g of gluten extract showing normal glomeruli (b), kidneys of rats fed 1.0 g of gluten extract showing normal glomeruli (c), kidneys of rats fed 1.5 g of gluten extract showing normal glomeruli (d), livers of control rats showing normal hepatocytes (e), liver of rats fed 0.5 g of gluten extract showing normal hepatocytes (f), livers of rats fed 1.0 g of gluten extract showing normal hepatocytes (g), livers of rats fed 1.5 g of gluten extract showing normal hepatocytes (h).

## 4. Discussion

The non-significant body weight change of the treated rats compared to the control rats suggests that gluten may not have an adverse effect on body weight. Gluten is a protein compound that is low in carbohydrates, and documented evidence suggests that low-carbohydrate diets can reduce fat mass for up to six months (Wylie-Rosett *et al.*, 2013; Barber *et al.*, 2021). The rats in the current study were treated for three months, a duration that falls within the stated period during which low-carbohydrate diets are effective against weight gain. Consistent with Silva *et al.* (2020), our findings align with their cross-sectional study, which demonstrated no impact of gluten on body weight. Metwally and Farahat (2015) also found no difference in body weight between birds fed different gluten diets and the control group. However, the result contrasts with that of Freire *et al.* (2016), who found that gluten increases adiposity and, as a result, body weight in some gluten-fed animals. Furthermore, You *et al.* (2020) discovered a link between gluten consumption and obesity in a cross-sectional study involving 28 countries from both developed and developing countries. After 6 months, the efficacy of low-carbohydrate diets in maintaining body weight declines, and gluten may contribute to body weight gain by disrupting metabolism and providing excess energy (Barber *et al.*, 2020; You *et al.*, 2020). Therefore, studies that report weight gain may have a longer duration. It is important to consider additional factors that have been highlighted in relevant studies. For instance, the study by Jonson *et al.* (2015) indicates that gluten inhibits the binding of leptin to its receptor, which could potentially impact body weight.

Furthermore, Zong *et al.* (2020) suggest that gluten-containing foods often have a high carbohydrate content, which can also influence weight. Therefore, it is crucial to take into account these facts and their potential effects when examining the relationship between gluten consumption and body weight.

The non-significant differences in blood glucose levels between the control and treated rats further demonstrate gluten's non-toxicity. Gluten prolamin fraction regulates blood glucose by enhancing insulin synthesis or sensitivity in various organs (Nishizawa *et al.*, 2009). It also raises the expression of the adiponectin gene and thus its concentration, which prevents the accumulation of glucose in the blood by increasing fat metabolism in tissues (Nishizawa *et al.*, 2009). The findings of the current study are consistent with Hansen *et al.* (2018), who found no variations in blood glucose levels between gluten-free and gluten-containing diets. In addition, Shetty *et al.* (2020) found no influence on blood glucose levels in mice fed a gluten-containing diet. Furthermore, Metwally and Farahat (2015) found no significant changes in blood glucose levels in birds fed on maize, wheat, or rice gluten. In a meta-analysis of gluten's effects, those who consumed the most gluten-containing foods had reduced incidences of type 2 diabetes and mortality (Mellen *et al.*, 2008). This implies that gluten has no negative impact on blood glucose levels.

Although the lymphocytes (a hematological parameter) and ALP, ALT, and TP (liver enzymes) of the treated rats were significantly altered, their values remained within normal limits. Generally, a low-calorie diet like that of a gluten-rich diet may cause transient changes in the liver enzymes observed in the current study (Gasteyger *et al.*, 2008). The increase observed in the lymphocytes of the treated rats suggests immune cells' reaction to an infection or inflammation (Hamad and Mangla, 2021). Gluten may act as a probiotic, according to Neyrinck *et al.* (2012) and Tojo *et al.* (2014), promoting the growth of beneficial gut bacteria such as bifidobacteria. This suggests that the treated rats' elevated lymphocyte numbers could represent a reaction to the increased gut bacteria. Protein levels that were elevated in the treated rats could indicate malnutrition or mineral deficiencies (NHS, 2018). Malnutrition can also be indicated by a rise in ALT and a reduction in ALP (Ray *et al.*, 2017; Karajibani *et al.*, 2021). This suggests that gluten consumption may result in mineral or nutrient malabsorption. However, a study is needed to ascertain this claim. The current study's findings are consistent with those of Yahaya *et al.* (2020a), who reported insignificant changes in hematological parameters and changes in liver enzymes within normal ranges in some gluten-fed rats. Metwally and Farahat (2015) also reported changes within normal ranges of some liver enzymes, total protein, albumin, and globulin in birds fed on rice, corn, or wheat gluten diets. Furthermore, Dinani *et al.* (2020) found that rice gluten had no negative effect on the hemato-biochemical profile of broiler chicks.

The micronuclei test revealed that both the control and treated rats had some micronuclei abnormalities, with no significant difference between both groups of rats. This shows that gluten consumption was not the cause of the cytogenetic abnormalities identified in the treated groups. Unfortunately, there is a paucity of documented studies available for comparison to the current study in terms of cytogenetic effects. Meanwhile, the histopathological examinations of the rats further demonstrate that gluten may be harmless, as the treated rats' kidneys and livers exhibit no histological alterations when compared to the control rats. Gluten is a protein that also contains some quantities of vitamins and minerals that protect and repair tissues (Kumar, 2011; Rosell, 2012). This finding is consistent with that of Yahaya *et al.* (2020a), who found no histological changes in the livers and kidneys of rats fed gluten-containing diets. Lebwohl *et al.* (2017) also reported that long-term consumption of gluten by some adult humans did not induce any cardiac tissue damage. Also, neither Viljamaa *et al.* (2005) nor Guidetti *et al.* (2001) found a link between a gluten diet and a higher risk of autoimmune diseases.

## 5. Conclusion and Recommendations

The results of this study have demonstrtaed that gluten has no significant adverse effects on body weight, blood glucose levels, hematological, hepatic, and renal function parameters, as well as the histology of rats. The lymphocytes, alkaline phosphatase, alanine transaminase, total protein, and calcium ions of the test rats were all altered ($P < 0.05$) but remained within the normal ranges. This implies that gluten-rich diets may not have toxic effects on human systems. However, the altered proteins, alkaline phosphatase, and alanine transaminase are signs of mineral or nutrient deficiency. This suggests that long-term gluten consumption may result in mineral and vitamin malabsorption; however, further studies are required to verify this assertion. The results also imply that prolonged consumption of gluten-containing diets without breaks should be avoided.

## 6. Acknowledgements

We would like to acknowledge the contributions of the staff of the Department of Biological Sciences, Federal University Birnin Kebbi, Nigeria.

## Ethical Statement

This study was conducted in accordance with the ethical standards of the European and German Animal Welfare legislation, declaration principles set out by Helsinki and the National Institutes of Health guidelines for care and use of animals in research. All protocols were approved by the local ethics committee of the Federal University Birnin Kebbi, Nigeria (regulation CEE 86/609).

Supplementary Table 1. Hematological parameters of rats treated with gluten extracts.

| Groups | PCV | Hb | WBC | RBC | Lym |
|---|---|---|---|---|---|
| Control | 0.26±0.01[a] | 11.22±0.11[c] | 14.23±0.02[b] | 5.20±0.04[d] | 63.32±0.03[d] |
| 0.5 | 0.27±0.03[a] | 12.87±0.05[a] | 14.57±0.04[a] | 5.76±0.02[a] | 71.62±0.04[c] |
| 1 | 0.27±0.01[a] | 12.08±0.04[b] | 14.00±0.04[c] | 5.63±0.02[b] | 72.98±0.01[a] |
| 1.5 | 0.26±0.03[a] | 11.15±0.02[c] | 13.72±0.03[d] | 5.57±0.04[c] | 72.18±0.04[b] |
| Mean | 0.27±0.02 | 11.83±0.74 | 14.13±0.33 | 5.54±0.22 | 70.03±4.07 |
| Mean square | 0.000 | 1.979 | 0.389 | 0.173 | 60.877 |
| p Value | 0.894[ns] | <0.001[*] | <0.001[*] | <0.001[*] | <0.001[*] |

Supplementary Table 2. Body weight parameters of rats treated with gluten extracts.

| Groups | Initial body weight | Final body weight | Change in body weight | p value |
|---|---|---|---|---|
| Control | 199.33±0.03 | 207.41±0.20 | −82.323 | <0.001[*] |
| 0.5 | 198.33±0.34 | 210.51±0.20 | −39.067 | 0.001[*] |
| 1 | 197.00±0.45 | 209.32±0.19 | −82.073 | <0.001[*] |
| 1.5 | 198.15±0.23 | 208.43±0.19 | −445.137 | <0.001[*] |
| Mean square | 2.736 | 5.208 | | |
| Mean | 198.20±0.90 | 208.92±1.20 | | |

Supplementary Table 3. Blood glucose levels in rats treated with gluten extracts.

| Groups | Initial blood glucose | Final blood glucose | t | p value |
|---|---|---|---|---|
| Control | 5.05±0.03 | 5.02±0.05 | 0.65 | 0.583[ns] |
| 0.5 | 5.22±0.04 | 5.13±0.03 | 2.227 | 0.156[ns] |
| 1 | 5.17±0.04 | 5.22±0.01 | -2.887 | 0.102[ns] |
| 1.5 | 5.22±0.01 | 5.14±0.05 | 3.464 | 0.074[ns] |
| Mean square | 0.019 | 0.02 | | |
| Mean | 5.17±0.08 | 5.13±0.08 | | |

Supplementary Table 4. Hepatic parameters of rats treated with gluten extracts.

| Groups | ALP | AST | ALT | TP | ALB |
|---|---|---|---|---|---|
| Control | 21.00±0.20[a] | 11.71±0.01[c] | 13.00±0.02[a] | 7.00±0.04[d] | 4.07±0.01[b] |
| 0.5 | 17.67±0.02[b] | 12.69±0.02[a] | 7.67±0.02[b] | 8.43±0.04[a] | 4.13±0.01[a] |
| 1 | 15.33±0.01[d] | 11.65±0.02[d] | 7.33±0.02[d] | 7.80±0.02[c] | 4.06±0.03[b] |
| 1.5 | 15.81±0.02[c] | 11.81±0.03[b] | 7.45±0.00[c] | 8.02±0.02[b] | 4.10±0.03[ab] |
| Mean | 17.45±2.33 | 11.97±0.44 | 8.86±2.50 | 7.81±0.54 | 4.09±0.03 |
| Mean square | 19.835 | 0.714 | 22.885 | 1.085 | 0.003 |
| p Value | <0.001[*] | <0.001[*] | <0.001[*] | <0.001[*] | 0.019[*] |

Supplementary Table 5. Renal function parameters of rats treated with gluten extracts.

| Groups | Na | K | Cl | Urea | Ca |
|---|---|---|---|---|---|
| Control | 140.17±0.01[b] | 4.15±0.03[a] | 102.14±0.04[c] | 3.05±0.01[a] | 76.50±0.40[b] |
| 0.5 | 140.17±0.02[b] | 4.05±0.04[b] | 102.83±0.02[b] | 3.03±0.02[a] | 71.67±0.01[c] |
| 1 | 136.17±0.01[c] | 4.05±0.03[b] | 103.03±0.06[a] | 2.93±0.03[b] | 82.50±0.30[a] |
| 1.5 | 142.33±0.06[a] | 4.07±0.04[b] | 102.83±0.04[b] | 3.02±0.03[a] | 66.50±0.20[d] |
| Mean | 139.71±2.33 | 4.08±0.05 | 102.71±0.35 | 3.01±0.05 | 74.29±6.18 |
| Mean square | 19.819 | 0.007 | 0.456 | 0.008 | 139.837 |
| p Value | <0.001[*] | 0.025[*] | <0.001[*] | 0.001[*] | <0.001[*] |

Supplementary Table 6. Micronuclei and nuclear aberrations in the rats treated with gluten extracts.

| Groups | Micronucleus | Nuclear aberration |
|---|---|---|
| Control | 5.33±0.01[c] | 6.50±0.40[a] |
| 0.5 | 5.33±0.01[c] | 6.44±0.02[a] |
| 1 | 6.67±0.04[a] | 6.33±0.04[a] |
| 1.5 | 6.33±0.01[b] | 6.40±0.10[a] |
| Mean | 5.92±0.62 | 6.42±0.19 |
| Mean square | 1.427 | 0.015 |
| p Value | <0.001[*] | 0.787[ns] |